\documentclass[a4paper]{jpconf}
\usepackage{graphicx}
\usepackage[utf8x]{inputenc}
\usepackage{caption}
\usepackage{adjustbox}
\usepackage{hyperref}
\usepackage{cite}

\begin{document}
\title{Dark matter search with the SABRE experiment}

\author{Giulia D'Imperio, for the SABRE collaboration}

\address{Physics Department of Sapienza Università di Roma, Piazza Aldo Moro, 2, 00185 Roma, Italy}

\ead{giulia.dimperio@roma1.infn.it}

\begin{abstract}
The SABRE (Sodium Iodide with Active Background REjection) experiment will search for an annually modulating signal from dark matter using an array of ultra-pure NaI(Tl) detectors surrounded by an active scintillator veto to further reduce the background. 
The first phase of the experiment is the SABRE Proof of Principle (PoP), a single 5~kg crystal detector operated in a liquid scintillator filled vessel at Laboratori Nazionali del Gran Sasso (LNGS). The SABRE-PoP installation is underway with the goal of running in 2018 and performing the first in situ measurement of the crystal background, testing the veto efficiency, and validating the SABRE concept.
The second phase of SABRE will be twin arrays of NaI(Tl) detectors operating at LNGS and at the Stawell Underground Physics Laboratory (SUPL) in Australia. By locating detectors in both hemispheres, SABRE will minimize seasonal systematic effects. 
This paper presents the status report of the SABRE activities as well as the results from the most recent Monte Carlo simulation and the expected sensitivity.

\end{abstract}

\section{Introduction}

According to the standard model of cosmology, the baryonic matter constitutes only about 5\% of the mass-energy content of the universe; about 25\% is made of  ``dark matter", and the remaining is ``dark energy".
The nature of dark matter (DM) is unknown but a family of well motivated candidates are the WIMPs (weakly interacting massive particles).

A worldwide effort has been spent in the last years to measure directly the WIMP interaction with different target materials.   
The only experiment observing a signal compatible with the DM hypothesis is DAMA/LIBRA, an array of 250~kg NaI(Tl)  crystals operating at Laboratori Nazionali del Gran Sasso (LNGS). 
DAMA/LIBRA  measures the annual modulation in the detector rate due to the motion of the Earth through the hypothetical WIMP halo.
The modulation observed 
in single-hit rate in the  2--6 keV energy region
 has high significance (9.3~$\sigma$, combining the data of the two phases, DAMA/LIBRA and DAMA/NaI~\cite{Bernabei:2008yi,Bernabei:2013xsa}), but the DM interpretation is controversial, since other experiments (LUX\cite{Akerib:2013tjd}, XENON~\cite{Aprile:2012nq}, CDMS~\cite{Agnese:2015ywx}),  give null results in the WIMP parameters phase space compatible with DAMA signal. 
An independent measurement using NaI(Tl) detectors is therefore required.

\section{The SABRE experiment}
\label{sec:strategy} 

SABRE is a new experiment aiming at the DM search through annual modulation with ultra-radiopure NaI(Tl) detectors. 
The SABRE strategy to reach high sensitivity and test the DAMA result relies on:
\begin{itemize}
	\item background suppression of both external and internal sources. The goal is to reach a total background lower than DAMA's 1 cpd/kg/keV. 
		This is achieved using ultra-radiopure crystals and using an active veto system in addition to the passive shielding;
	\item low energy threshold (equal or lower than DAMA's 2 keV). This is obtained using high quantum efficiency photomultipliers directly coupled to crystals;
	\item double location in North and South hemisphere, in order to disentangle seasonal effects.
\end{itemize}

The SABRE experiment will have two phases.
The first is the ``Proof of Principle" (PoP) which  is currently being deployed at LNGS and will take data in 2018.
The scope of the SABRE-PoP is to validate the general strategy, and have a first measurement of the crystal background and of the veto efficiency. 
The SABRE-PoP setup is therefore simplified with respect to the full scale, and the detector will consist in one single NaI(Tl) crystal of 5~kg. 

The second phase is the full scale experiment, which will consist in twin detectors of at least $\sim$50~kg, located one in the North and one in the South hemisphere, respectively at Laboratori Nazionali del Gran Sasso (Italy) and Stawell Underground Physics Laboratory (Victoria, Australia).

The SABRE collaboration includes $\sim$50 people from 11 institutions in Italy (LNGS, INFN Roma 1, INFN Milano), US (Princeton University, Lawrence Livermore National Laboratory (LLNL), Pacific Northwest National Laboratory (PNNL)), UK (Imperial College), and Australia (Australian National University, University of Adelaide, University of Melbourne, Swinburne University of Technology).

\subsection*{SABRE-PoP design}
\label{sec:PoPdesign}

From the outside to the inside, the SABRE Proof of Principle setup is composed of the following parts:
\begin{itemize}
	\item a passive shielding made  of an external layer of water tanks on top (80~cm) and sides (90~cm), polyethylene (PE) walls (40~cm), door (65~cm) top (10~cm) and floor (10~cm). An additional steel plate of 2~cm is placed on the top, between the PE and the water, and 15~cm of lead are placed on the floor, under the PE base. 
The internal volume of the shielding will be flushed with nitrogen in order to remove the environmental radon background;
\item a steel vessel of 1.3~m diameter  and 1.5~m length. The internal surface is covered with lumirror polyester film in order to have high reflectance and improve light collection;
\item 2 tons of liquid scintillator (PC + 3~g/l of PPO) filling the steel vessel and acting as a veto system.   
The scintillation light is read by 10 8" Hamamatsu R5912 PMTs with high quantum efficiency;
\item the crystal insertion system (CIS) to keep the crystal module isolated from the liquid scintillator. It consists of a tube of radiopure copper, and a steel bar connected to the crystal module. Also the volume inside the CIS tube will be flushed with nitrogen to prevent radon contamination;
\item the crystal module made of a copper enclosure, containing one NaI(Tl) crystal of 5~kg weight and  diameter 4", 
	directly coupled to two PMTs of 3" Hamamatsu R11065-20 with high QE and high radiopurity. 
\end{itemize}

\begin{figure}[!h] 
    \centering
    \includegraphics[width=0.33\textwidth]{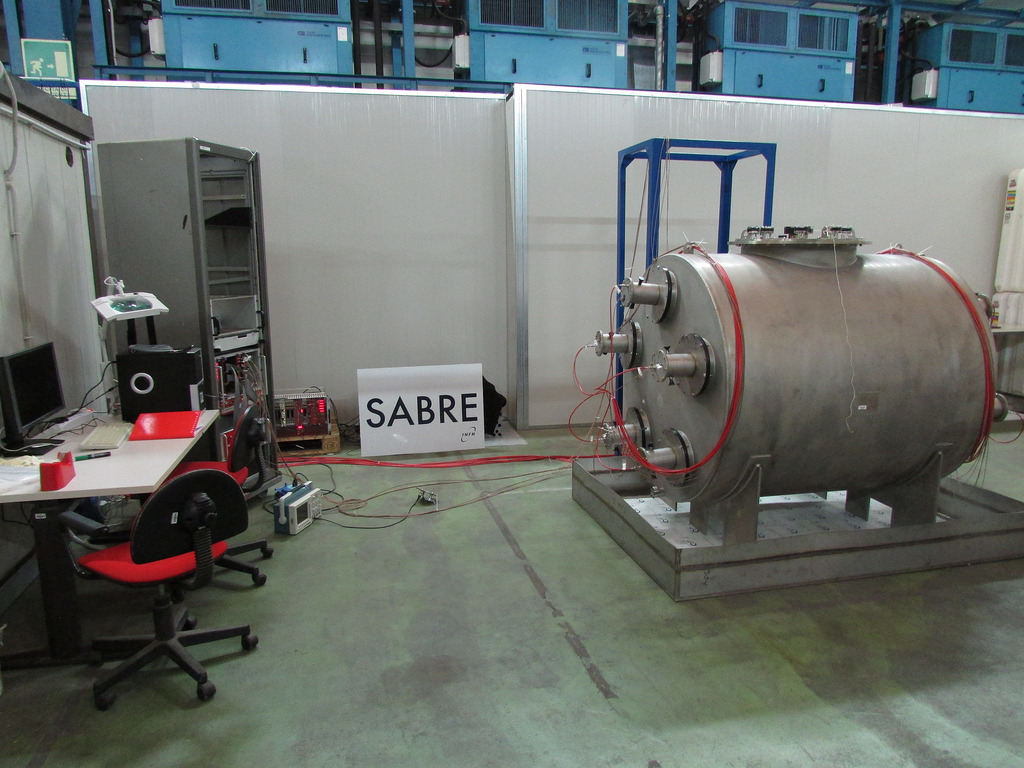} \qquad
    \includegraphics[width=0.33\textwidth]{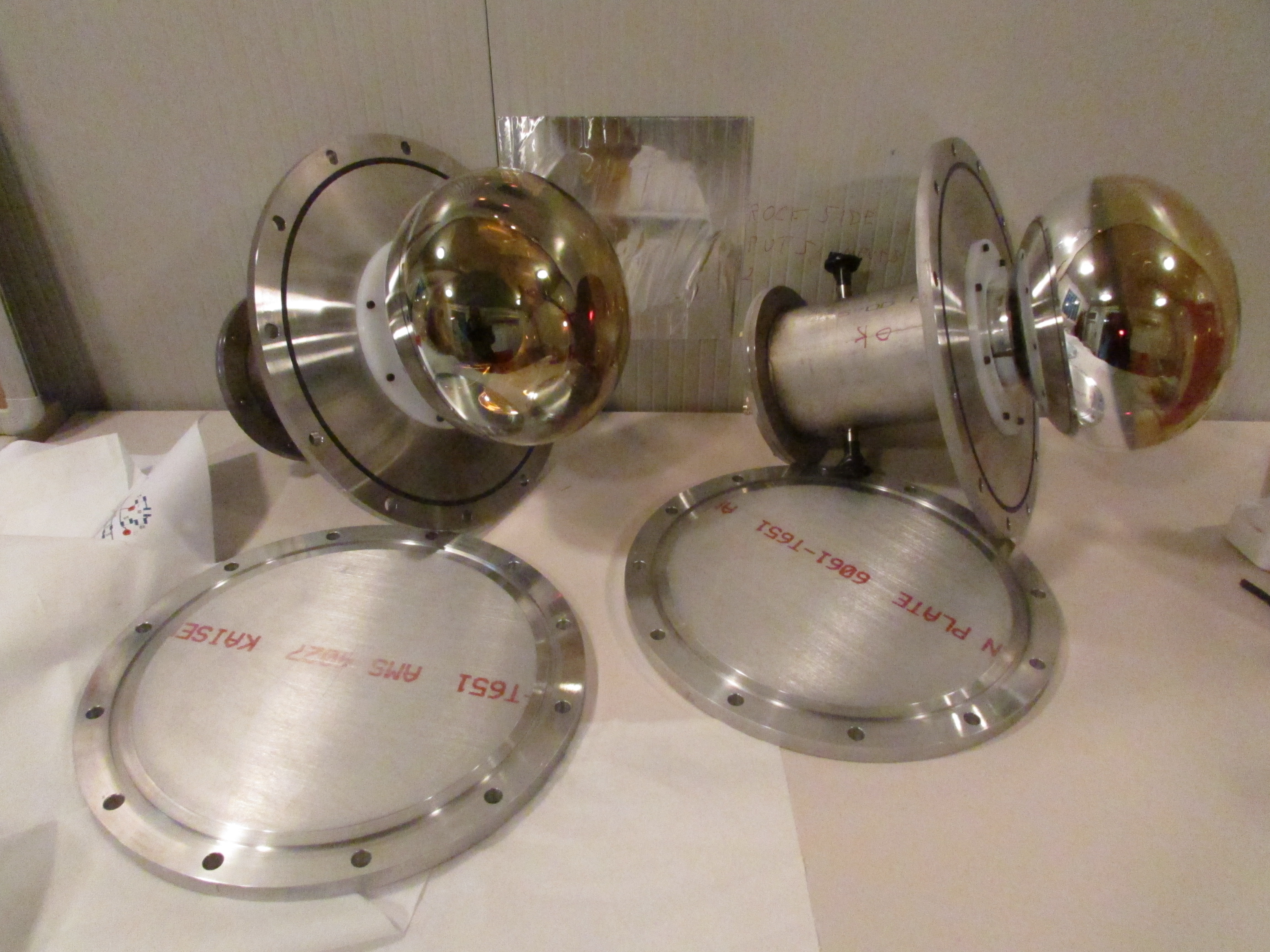}
	\caption{\label{fig:SABRE-PoP}Steel vessel of SABRE-PoP installed in Hall B at LNGS (left) and 8" PMTs Hamamatsu R5912 mounted on the vessel flanges (right).
}
\end{figure}

The steel vessel with its 10 PMTs, which will contain the liquid scintillator (LS) veto, is installed at LNGS in a preliminary site in Hall B (figure \ref{fig:SABRE-PoP}).
The final site for the SABRE-PoP will be in Hall C, where the refurbishment of the floor has been recently completed (May 2017) and the installation of the passive shielding has started in June.   

The SABRE crystals will be grown at RMD (Boston) with NaI Astro Grade powder from Sigma Aldrich.
The contamination of relevant isotopes in Astro Grade powder is 9 ppb for K, $<$0.1 ppb for Rb, $<$1 ppt for U, and $<$1 ppt for  Th. 
A 2~kg crystal of nearly the final diameter has been produced in 2015 and the K concentration resulted to be the same as Astro Grade powder (9 ppb)~\cite{DAngelo:2017lxm}.
The final 5~kg crystal for the SABRE-PoP  is currently in production.

\section{Simulation of the SABRE-PoP background}
\label{sec:simulation}

The contributions to the background from intrinsic radioactivity and cosmogenic activation of the crystals and surrounding materials have been studied with new Monte Carlo simulations that reproduce carefully the geometrical volumes of the experimental setup. 
All internal backgrounds have been evaluated up to the external shielding (excluded), which is implemented in the geometry.
The instrinsic contamination values for the crystal used in the Monte Carlo are those measured for the 2~kg test crystal produced from Astro grade powder~\cite{DAngelo:2017lxm}. For cosmogenics and other materials, contamination values have been taken from literature~\cite{Aprile:2015lha,Amare:2016rbf,Bernabei:2008yh,Alduino:2016vtd,Baudis:2015kqa,Agnes:2015qyz,Alimonti:2009zz}.
The results are in agreement with previous simulations~\cite{Froborg:2016ova,DAngelo:2017lxm}.

\subsection*{Potassium measurement mode}
One of the first goals of the SABRE-PoP is to measure the $^{40}$K content in the crystal, which is one of the most significant background contributions to the energy spectrum in the region of interest. 

A  $^{40}$K decay in 11\% of cases gives an energy release around 3 keV due to X-rays or Auger electrons in coincidence with a $\gamma$ of 1.46 MeV. 
The $\gamma$ can escape from the crystal volume and release part of its energy in the LS.
An event in potassium measurement mode (KMM) is a potassium-like event, defined as an energy deposition between 2 and 4 keV in the crystal (1~$\sigma$ around the 3 keV $^{40}$K peak) in coincidence with an energy deposit between 1.28 MeV and 1.64 MeV in LS (2.5~$\sigma$ around the 1.46 MeV $^{40}$K peak). 
True coincidences from  background sources other than $^{40}$K in the crystals can mimic the KMM signature and have therefore to be included.

\begin{figure}[!h]
\begin{minipage}{\textwidth}
  \begin{minipage}[b]{0.5\textwidth}
    \centering
        \includegraphics[width=0.7\textwidth]{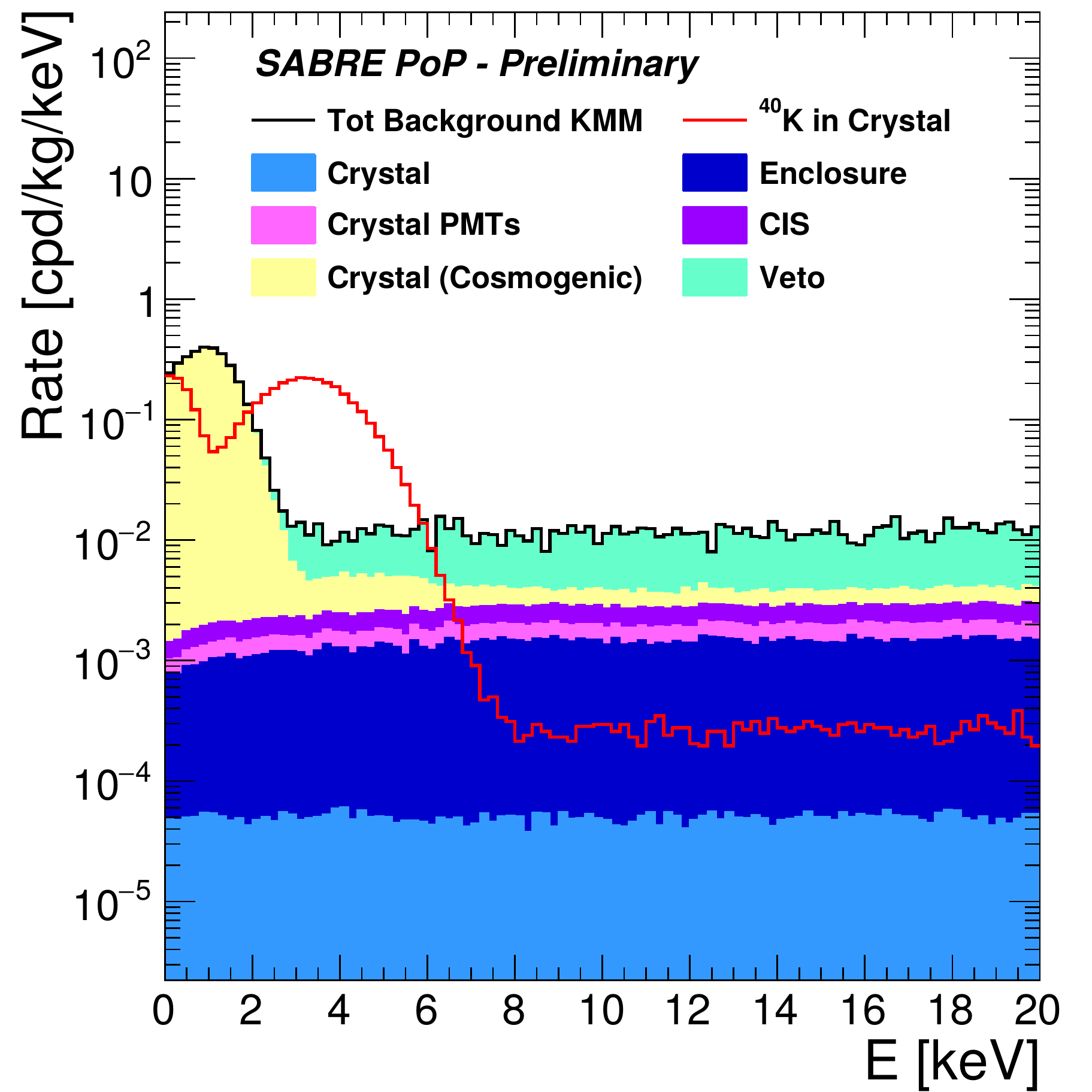}
    \captionsetup{width=0.9\linewidth}
    \captionof{figure}{\label{fig:KMM60_lowEne}
	  Background from all SABRE-PoP setup components in a $(0-20)$~keV region in potassium measurement mode (KMM). 
\\}
  \end{minipage}
  \hfill
  \begin{minipage}[b]{0.53\textwidth}
    \centering
        \footnotesize
      \begin{tabular}{cc}
      \br
              &  Rate KMM    \\
              &  [cpd/kg/keV]\\
      \mr
        Crystal                         &     $5.1 \cdot 10^{-5}$\\
        Crystal Cosmogenic*             &     $1.8 \cdot 10^{-2}$\\
        Crystal PMTs                     &     $3.7 \cdot 10^{-4}$\\
        Enclosure*                      &     $1.3 \cdot 10^{-3}$\\
        CIS*                            &     $7.7 \cdot 10^{-4}$\\
        Veto                            &     $6.2 \cdot 10^{-3}$\\  
      \mr
        Total                           &     $2.7 \cdot 10^{-2}$\\
        \textbf{Crystal $^{40}$K}       &     \textbf{1.9 $\cdot$ 10$^{-1}$}\\
      \br
      \end{tabular}
	  \captionof{table}{\label{bulk_kmm_total}Background rates in the region $(2-4)$~keV in KMM from all the SABRE-PoP setup component. 
      *Cosmogenic backgrounds are computed after 60 days underground.\\ \\
      }
        
  \end{minipage}
\end{minipage}
\end{figure}

The  expected backgrounds in KMM and the K rate for a 10 ppb contamination are shown in figure~\ref{fig:KMM60_lowEne}: the contributions from each setup component are stacked and the total is the solid black line. 
The line labelled ``Crystal" does not include the K component, which is superimposed in red and represents the signal in this configuration.
``Veto" includes the steel vessel, the LS and the 10 PMTs. 
Expected rates in the crystal in the (2--4)~keV interval are  summarized in table~\ref{bulk_kmm_total}.
From these results (assuming a K content in the crystal at the level of 10 ppb) we expect to reach 1 ppb precision on the K contamination measurement in about two months of data taking.

\subsection*{Background for dark matter measurement}

For the dark matter measurement (DMM) we are interested in the background rate between 2 and 6 keV in the crystal (the DAMA modulation region). 
The liquid scintillator is used as an active veto to reduce K and other intrinsic and cosmogenic backgrounds.
The veto threshold is set to 100 keV with $\sim$100\% efficiency on the light collection. 

\begin{figure}[!h]
\begin{minipage}{\textwidth}
  \begin{minipage}[b]{0.5\textwidth}
    \centering
        \includegraphics[width=0.7\textwidth]{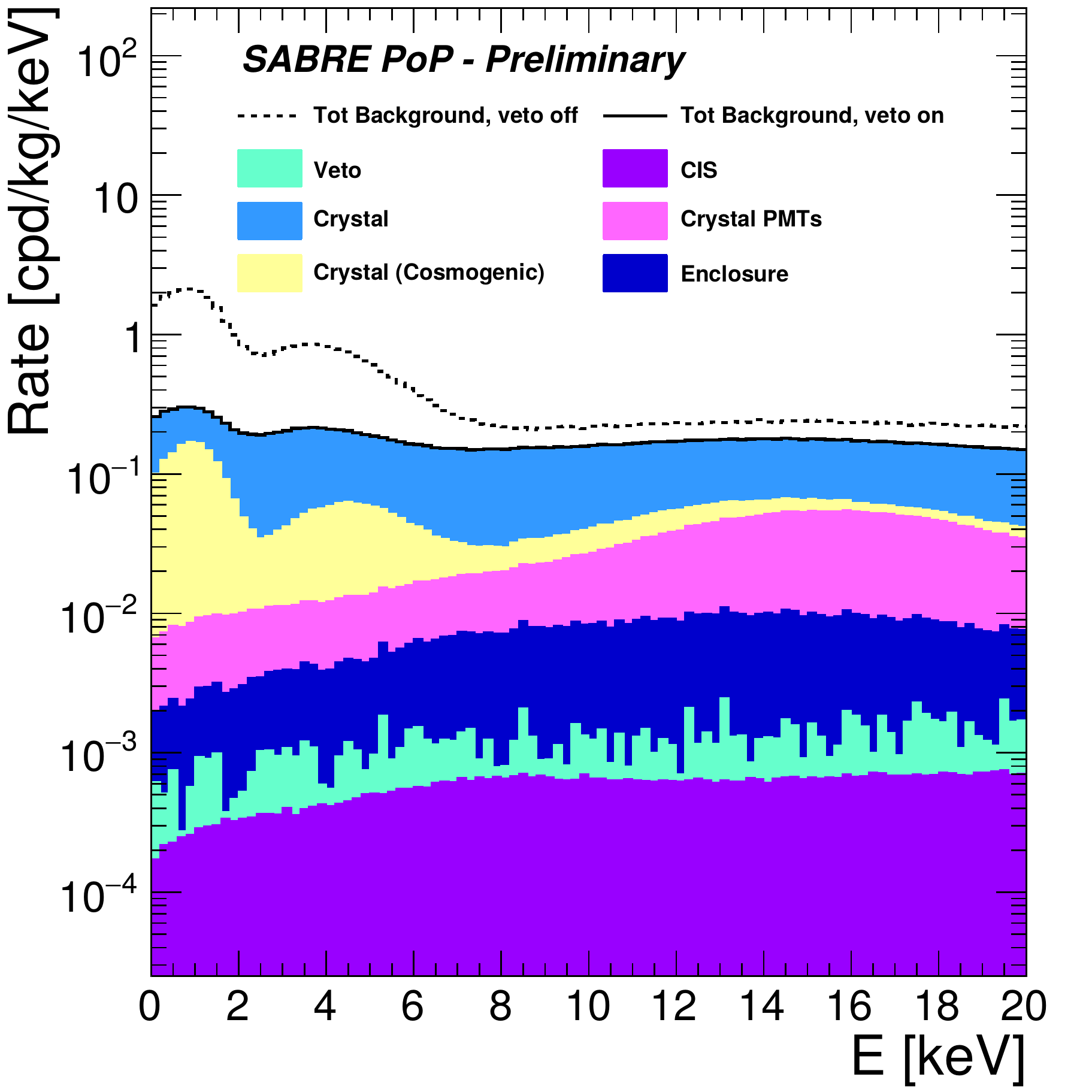}
    \captionsetup{width=0.9\linewidth}
    \captionof{figure}{\label{fig:DMM180_lowEne}
	  Backgrounds from the SABRE-PoP setup components in a $(0-20)$~keV region in dark matter measurement mode (DMM).\\}
  \end{minipage}
  \hfill
  \begin{minipage}[b]{0.53\textwidth}
    \centering
        \footnotesize
    \begin{tabular}{ccc}
    \br
            & Rate, veto OFF & Rate, veto ON\\
            & [cpd/kg/keV]   & [cpd/kg/keV]\\
    \mr
    
      Crystal               &  $3.5 \cdot 10^{-1}$ &    $1.5 \cdot 10^{-1}$\\
      Crystal Cosmogenic* &  $3.0 \cdot 10^{-1}$ &    $3.9 \cdot 10^{-2}$\\
      Crystal PMTs           &  $1.2 \cdot 10^{-2}$ &    $8.1 \cdot 10^{-3}$\\
      Enclosure*          &  $9.5 \cdot 10^{-3}$ &    $3.6 \cdot 10^{-3}$\\
      CIS*                &  $3.7 \cdot 10^{-3}$ &    $4.6 \cdot 10^{-4}$\\
      Veto                  &  $3.0 \cdot 10^{-2}$ &    $5.7 \cdot 10^{-4}$\\  
    
    \mr
      Total                 &  $7.1 \cdot 10^{-1}$ &    $2.0 \cdot 10^{-1}$\\
    \br
    \end{tabular}
     \captionof{table}{\label{bulk_total}
	  Background rate in the region of interest $(2-6)$~keV from the SABRE-PoP setup components, with veto off and on respectively.\\ 
    **Cosmogenic backgrounds are computed after 180 days underground.\\ \\}
  
  \end{minipage}
\end{minipage}
\end{figure}

Crystal contamination gives the most relevant contribution to the background: it gives a rate of  $1.5 \cdot 10^{-1}$ cpd/kg/keV on a total of $2.0 \cdot 10^{-1}$ cpd/kg/keV (table~\ref{bulk_total}).
It must be noted however that, apart from the $^{40}$K contribution, the contamination levels in the crystals correspond to upper limits.
Figure~\ref{fig:DMM180_lowEne} shows the stacked backgrounds from the various setup components. 
The total with veto on is shown with a solid black line; the dashed black line indicates the total background with veto off.

The veto reduces the total internal backgrounds in the region of interest (2--6) keV by a factor 3.5 and has a rejection efficiency of 84~\% on the K in the crystal.

Backgrounds from the shielding materials and from external gammas are not included in figure~\ref{fig:DMM180_lowEne} and table~\ref{bulk_total}, but 
a Monte Carlo estimation has been done for the polyethylene  of the shielding and for the gamma external background.
For both sources we expect less than 10$^{-3}$ cpd/kg/keV.
These numbers include the veto effect, which is about a factor 100 reduction.

\section{Expected sensitivity}
\label{sec:sensitivity}

We have investigated the SABRE sensitivity to the annual modulation in the (2--6) keV$_{ee}$ energy region assuming the Standard Halo Model~\cite{Savage:2008er} and two different background hypotheses: the background estimated by Monte Carlo simulations for SABRE of 0.2 cpd/kg/keV and a higher background of 1~cpd/kg/keV.

\begin{figure} [!ht]
\centering
\includegraphics[width=0.54\textwidth]{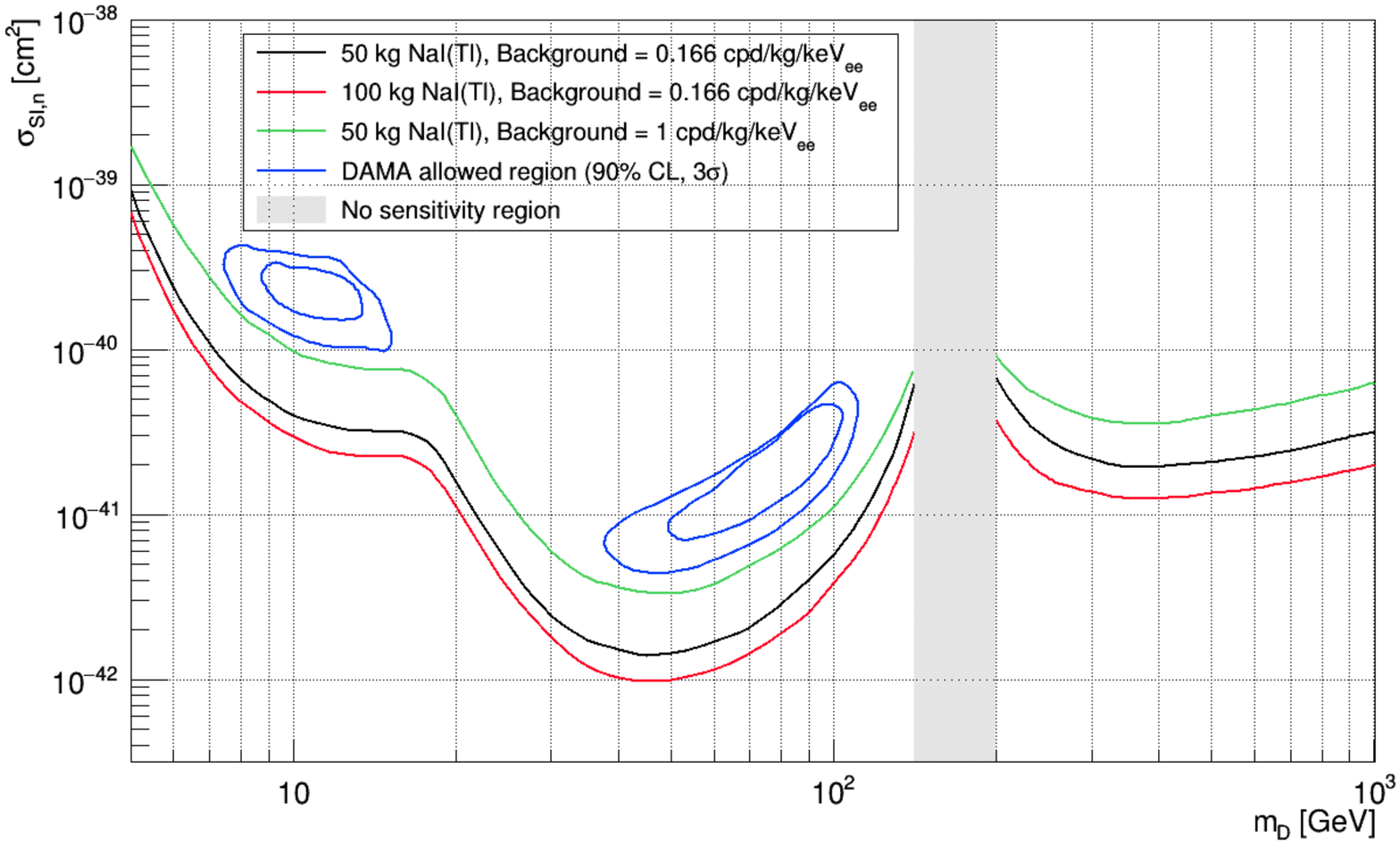}
\caption{Sensitivity plots for a 3-year data taking period and for the recoil energy range $(2-6)$~keV$_{ee}$. 
The black, red and green curves show the 90\%CL upper limit using different background hypotheses  and detector masses.
The blue curves show the DAMA allowed region from Savage et al. \cite{Savage:2008er}. 
The ``no sensitivity region" correspond to the WIMP mass region where the DM rates at maximum and minimum Earth relative velocity cross each other in the $(2-6)$~keV$_{ee}$ interval, and therefore the modulation amplitude goes to 0.
	}
\label{fig:Sensitivity_plots}
\end{figure}

Expected limits at 90\%~CL have been calculated for the two backgrounds and for two mass hypotheses (50 and 100 kg). 
The limits are shown in figure~\ref{fig:Sensitivity_plots}.

This study indicates that in three years of data taking SABRE will be sensitive to both the WIMP parameters regions obtained from fit to DAMA's results in the standard WIMP scenario~\cite{Freese:2012xd}.
From figure~\ref{fig:Sensitivity_plots} is also evident that we gain more in sensitivity  by lowering the background from 1 to 0.2~cpd/keg/keV than by doubling the detector mass.

\end{document}